# Formation of primordial helium: a rapid and simple scheme of calculation

Tamaz Kereselidze[1], Zaal Machavariani[1,2] and Irakli Noselidze[3]

[1]Faculty of Exact and Natural Sciences, Tbilisi State University, Chavchavadze Avenue 3, 0179 Tbilisi, Georgia
[2]Doctoral School, Kutaisi International University, Youth Avenue, 5th Lane, 4600 Kutaisi, Georgia
[3]School of Science and Technology, University of Georgia, Kostava Str. 77a, 0171 Tbilisi, Georgia

**ABSTRACT**

A rapid and relatively simple scheme of calculation is elaborated and applied to cosmological recombination of helium. Employing the nonrelativistic Coulomb Green's function, a wavefunction of a colliding electron is represented in an integral form applicable for calculations. Bound electrons of helium are described by the Hartree-Fock wavefunctions. The free-bound transition probabilities into excited states of helium, and the probabilities of bound-bound transitions in helium are calculated in different modes. It is revealed that free-bound transition probabilities weakly depend on to what extent a field experienced by a colliding electron deviates from the purely Coulomb field with charge $Z=1$, whereas these probabilities strongly depend on the choice of a wavefunction of a bound active electron involved in recombination.

1. **INTRODUCTION**

Around a few hundred thousand years after the beginning of the Universe its temperature reduced to a point at which formation of neutral helium and hydrogen atoms became energetically favoured. At the end of recombination era (redshift $z \simeq 1100$) the electromagnetic radiation effectively decoupled from matter and since then photons travelled without scattering throughout the Universe. Nowadays these photons constitute what is detected as the cosmic microwave background (CMB) radiation. The CMB spectrum conveys important imprints of the recombination era.

For an electron and proton, cosmological recombination was first studied by Zel'dovich, Kurt & Syuniaev (1968) and slightly later by Peebles (1968). According to these authors an electron and a proton combined efficiently into the hydrogen atom only in a highly excited state, from which a rapid cascade occurred into a state with principal quantum number $n = 2$. A radiative decay from state $2p$ involving one photon or from state $2s$ involving two photons then yielded the hydrogen atom in its ground state. The decay $2p \to 1s$ led to the appearance of photons with energy sufficient for the Lyman-$\alpha$ resonance excitation of atomic hydrogen already formed in the ground state. The $2p \to 1s$ decay processes were thus compensated by the excitation processes. The $2s \to 1s$ two-photon decay, which is about eight orders of magnitude slower than the $2p \to 1s$ one-photon decay, was thus the dominant process for the formation of hydrogen in the ground state. Since the probability of $2s \to 1s$ decay is small ($W_{2s,1s} = 8.227\, s^{-1}$), the population of the ground state of $H$ differs from the equilibrium one. As a result, the recombination is stretched in comparison with the equilibrium process taking place according to the Saha-Boltzmann law. The present level of investigation of the problem can be found in a review of Kurt & Shakhvorostova (2014) and in references therein.

In our recent paper (Kereselidze, Noselidze & Ogilvie 2019), we estimated the average distance between protons in the pre-recombination stage of evolution of the Universe. The estimation revealed that this distance was comparable with a linear size of the hydrogen atom being in a highly excited state. This means that the nearest neighbouring proton affected on the hydrogen atom, and hence participated in the recombination. The presence of another proton reduces the symmetry of a field experienced by an electron involved in recombination from spherical to axial and leads to a Stark splitting of the hydrogen energy levels. These two effects lead in turn to radiative transitions that are forbidden in the recombination of an electron with an isolated proton. Our subsequent calculations (Kereselide, Noselidze & Ogilvie 2021, 2022) showed that the non-standard mechanism of hydrogen recombination changes a thermal history of the Universe.

Because of larger ionization potential, $HeII$ recombined before $H$. It was assumed that as for $H$, the direct formation of $HeII$ in the ground state by recombination was not efficient, since $HeII$ were immediately ionized by the released energetic photons. More efficient was recombination through the intermediate state with $n=2$. The main mechanism of the formation of $HeII$ was thus two-photon decay of the 2s state. For $HeII$ the probability of $2s \to 1s$ decay is $W_{2s,1s} = 526.5\, s^{-1}$; the rate of this process is $N_{2s} W_{2s,1s}$, in which $N_{2s}$ is the population of the $2s$-level. Since $N_{2s} W_{2s \to 1s}$ is greater than the recombination rate of electrons in the $2s$-level, the electrons do not detain in this level and rapidly transit into the ground state. This assumption was supported by the numerical calculations (Seagel, Sasselov & Scott 2000), which showed that the $HeIII \to HeII$ recombination proceeds in fact according to the Saha-Boltzmann law. Obviously, an influence of the nearest neighboring ion $HeIII$ on the process will change the Saha-Boltzmann scenario of recombination.

The physics of $HeI$ recombination differs from that for $HeII$ and $H$ recombination because of its different atomic structure. Unlike early calculations where helium was treated as a three-level atom, modern numerical calculations of cosmological recombination use a multi-level atom model where the fine structure of levels is taken into account. This allows both the singlet and triplet states of helium to be taken into account in the calculations. It was shown that $HeI$ recombination takes place in a mode that is different from the Saha-Boltzmann equilibrium mode (Switzer & Hirata 2008, Wong, Moss & Scott 2008, Grin & 2010, Chluba, Thomas 2011).

The exact wavefunctions of two electrons being in the field of helium nucleus cannot be found analytically neither for the discrete nor the continuous spectra. This fact leads to additional difficulties in comparison with hydrogen recombination. Indeed, an application of numerical wavefunctions to $HeI$ recombination requires formidable computational efforts and is time consuming. The problem hence requires an alternative treatment.

In the present paper, we elaborate a rapid and relatively simple scheme of calculation applicable for $HeI$ recombination. Our approach is based on reducing the two-electron problem to the one-electron treatment. This allows us to find the wavefunctions of an active electron involved in recombination in a closed algebraic form. For this, we make use the nonrelativistic Coulomb Green's function (CGF) defined in an integral form. All these in turn make it possible to calculate transition probabilities without increasing the computation time significantly. Specifically, we calculate the probabilities of free-bound and bound-bound radiative transitions for $HeI$. Furthermore, we investigate the influence of a shielding of a nuclear charge by a bound electron on the transition probability. The influence of the nearest neighbouring ion on $HeI$ recombination is not taken into account; this will be treated in the separate paper.

The CGF can be constructed from its spectral representation, in which the summation runs over the complete set of discrete and continuum eigenstates (Baz, Perelomov & Zel'dovich 1969). Blinder (1981) showed that a summation explicitly written in term of discrete and continuous eigenstates in parabolic coordinates leads to the integral representation of the CGF. Making use of the scheme of calculation developed by Blinder, we evaluate the CGF in the form convenient for our purpose (equation (7) in Section 2).

The paper is organized as follows. After stating our objective, we derive the wavefunctions of an active electron in Section 2. In Section 3, we present the results of calculations and draw conclusions in section 4. Unless otherwise indicated, atomic units ($e = m_e = \hbar = 1$) are used throughout the paper.

## 2. WAVEFUNCTIONS OF ELECTRONS

A precise quantum-mechanical calculation of cosmological recombination requires a knowledge of the correct wavefunctions of electrons involved in the process in both the initial continuous and final discrete states. From a wave-mechanical point of view, the problem is to obtain the correct wavefunctions that are solutions of the Schrödinger equation,

$$\left(-\frac{1}{2}\Delta_1 - \frac{1}{2}\Delta_2 - \frac{2}{r_1} - \frac{2}{r_2} + \frac{1}{|\vec{r}_1 - \vec{r}_2|}\right)\Psi = E\Psi. \quad (1)$$

Here, $r_1$ and $r_2$ are the distances from electrons to the Coulomb centre and $E$ is the electron energy. Our purpose is to find the eigenfunctions of equation (1) in closed algebraic forms that are applicable to cosmological recombination.

2.1 Continuous spectrum wavefunction

The initial state wavefunction, we represent as this symmetrized function

$$\Psi_i^{(\pm)} = \frac{1}{\sqrt{2}}\left(\psi_{1s}(\vec{r}_1)\psi_{\vec{k}}(\vec{r}_2) \pm \psi_{1s}(\vec{r}_2)\psi_{\vec{k}}(\vec{r}_1)\right), \quad (2)$$

in which $\psi_{1s}$ is the wavefunction of a bound electron being in the ground state of $HeII$, and $\psi_{\vec{k}}$ is the wavefunction of a colliding electron. Signs (+) and (−) correspond to the total spin of electrons $S = 0$ (singlet states) and $S = 1$ (triplet states), respectively.

Inserting $\Psi_i$ into (1), multiplying on the left by $\psi_{1s}(\vec{r}_1)$ and integrating over $\vec{r}_1$, we obtain this equation for unknown wavefunction $\psi_{\vec{k}}^{(\pm)}(\vec{r})$

$$\left(-\frac{1}{2}\Delta - \frac{2}{r} + V(r) - \frac{k^2}{2}\right)\psi_{\vec{k}}^{(\pm)}(\vec{r}) = Q^{(\pm)}(\vec{r}). \quad (3)$$

In (3) $k^2/2 = (E - E_{1s})$ is the energy of a colliding electron and

$$V(r) = \frac{1}{r} - \left(2 + \frac{1}{r}\right)\exp(-4r), \quad (4)$$

$$Q^{(\pm)}(\vec{r}) = \mp\left\langle\psi_{1s}(\vec{r}\,')\left|\frac{1}{|\vec{r}-\vec{r}\,'|}\right|\psi_{\vec{k}}^{(0)}(\vec{r}\,')\right\rangle\psi_{1s}(\vec{r}). \quad (5)$$

When calculating $Q^{(\pm)}$, we replace $\psi_{\vec{k}}$ by $\psi_{\vec{k}}^{(0)}$ - the solution of equation (3) with $V = Q \equiv 0$ and assume that $\psi_{\vec{k}}^{(0)}$ and $\psi_{1s}$ are orthogonal functions. In equation (3) $V(r)$ describes the shielding of a nuclear charge by a bound electron, whereas $Q^{(\pm)}(\vec{r})$ accounts the influence of a colliding electron on a bound electron. Calculations show that the shielding effect much exceeds the impact of a colliding electron on a bound one (see appendix A). This fact allows us to neglect $Q^{(\pm)}(\vec{r})$ and include only $V(r)$ in the treatment. We thereby substantially reduce the time of calculations.

In the Coulomb-Born approximation the eigenfunction of equation (3) is expressible as

$$\psi_{\vec{k}}(\vec{r}) = \psi_{\vec{k}}^{(0)}(\vec{r}) + \int G^{(+)}(\vec{r},\vec{r}\,')\upsilon(r')\psi_{\vec{k}}^{(0)}(\vec{r}\,')d\vec{r}\,', \qquad (6)$$

in which $\upsilon = (2+1/r)\exp(-4r)$ and

$$G^{(+)}(\vec{r},\vec{r}\,') = -\frac{ik}{2\pi}\sum_{\bar{m}=-\infty}^{\infty} e^{i\bar{m}(\varphi-\varphi')} \int_0^{\infty} e^{i\frac{k}{2}(\mu+\nu+\mu'+\nu')\cosh s} \\ \times \chi(s) J_{\bar{m}}\left(b(\mu\mu')^{1/2}\right) J_{\bar{m}}\left(-b(\nu\nu')^{1/2}\right) ds, \qquad (7)$$

is the CGF. In (7) $\mu = r(1+\cos\vartheta)$, $\nu = r(1-\cos\vartheta)$, $\varphi = \arctan(y/x)$ are parabolic coordinates, in which $r$ is the radial variable, $\vartheta$ is the polar angle and $\varphi$ is the azimuthal angle; $\chi(s) = \sinh s \left(\coth(s/2)\right)^{2i/k}$, $b = k\sinh s$, and $J_{\bar{m}}(x)$ is a Bessel function. Symbol $(+)$ denotes an outgoing wave when $\vec{r} \to \infty$.

In (6) $\psi_{\vec{k}}^{(0)}$ is the solution of equation (3) with $V = Q^{(\pm)} \equiv 0$ and the nuclear charge $Z = 1$

$$\psi_{\vec{k}}^{(0)} = N e^{i\frac{k}{2}(\mu-\nu)} F(i/k, 1, ik\nu), \\ N = (2\pi)^{-3/2} e^{\pi/2k} \Gamma(1-i/k). \qquad (8)$$

Here, $F(i/k, 1, ik\nu)$ is a confluent hypergeometric function, $\Gamma(1-i/k)$ is a gamma function, and $N$ is a normalizing factor.

Inserting (7) into (6) and performing the integration over $\varphi'$, we obtain for $\psi_{\vec{k}}$ that

$$\psi_{\vec{k}} = \psi_{\vec{k}}^{(0)} - \frac{ik}{2}\int_0^{\infty} e^{i\frac{k}{2}(\mu+\nu)\cosh s} \chi(s) \int_0^{\infty}\int_0^{\infty} e^{-\frac{4-ik\cosh s}{2}\mu'} e^{-\frac{4-ik\cosh s}{2}\nu'} \\ \cdot J_0\left(b\sqrt{\mu\mu'}\right) J_0\left(-b\sqrt{\nu\nu'}\right) \psi_{\vec{k}}^{(0)}(\mu',\nu')(1+\mu'+\nu') d\mu' d\nu' ds. \qquad (9)$$

Here, the first term describes the motion of a colliding electron in the field of nucleus whose charge is fully shielded by a bound electron; the second term defines the correction caused by a partial shielding of a nuclear charge.

Inserting (8) into (9) and introducing the notations: $a_\mu = (4-ik(1+\cosh s))/2$, $a_\nu = (4+ik(1-\cosh s))/2$, we arrive at the following wavefunction for a colliding electron

$$\psi_{\vec{k}} = \psi_{\vec{k}}^{(0)} - \frac{ikN}{2}\int_0^{\infty} e^{i\frac{k}{2}(\mu+\nu)\cosh s} \chi(s)\left[A^{(0)}(s,\mu)B^{(0)}(s,\nu) \\ + A^{(1)}(s,\mu)B^{(0)}(s,\nu) + A^{(0)}(s,\mu)B^{(1)}(s,\nu)\right] ds. \qquad (10)$$

The functions $A^{(0)}$, $A^{(1)}$ and $B^{(0)}$, $B^{(1)}$ are defined and derived in an algebraic form in appendix B. We thus obtain that to determine the correction to $\psi_{\vec{k}}^{(0)}(\mu,\nu)$ only the integration over $s$ is required.

In Fig. 1 is shown the behaviour of $\left|\psi_{\vec{k}}(\mu,\nu)\right|^2$ when variable $\mu = 1$ and $\nu$ varies. In the same figure, we depict $\left|\psi_{\vec{k}}^{(0)}(\mu,\nu)\right|^2$ as a function of $\nu$ when $\mu = 1$. The energy of a colliding electron is assumed to be $1.06\, eV$ that corresponds to redshift $z = 3000$. Fig. 1 shows that both functions are oscillatory but with a decreasing amplitude as $\nu$ increases.

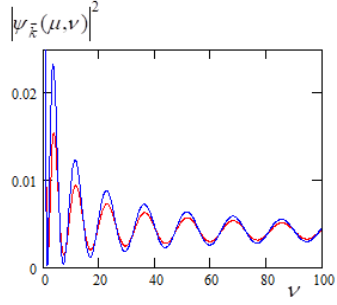

**Fig. 1.** The behaviour of $\left|\psi_{\vec{k}}(\mu,\nu)\right|^2$ (blue curve) and $\left|\psi_{\vec{k}}^{(0)}(\mu,\nu)\right|^2$ (red curve) when $\mu=1$ and $\nu$ varies.

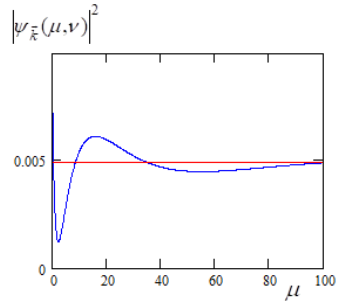

**Fig. 2.** As in Fig. 1 but when $\nu=1$ and $\mu$ varies.

When variable $\nu$ is fixed and $\mu$ varies, $\left|\psi_{\vec{k}}(\mu,\nu)\right|^2$ and $\left|\psi_{\vec{k}}^{(0)}(\mu,\nu)\right|^2$ are shown in Fig. 2. In this case $\left|\psi_{\vec{k}}^{(0)}(\mu,\nu)\right|^2$ does not depend on $\mu$ (see equation (8)), whereas $\left|\psi_{\vec{k}}(\mu,\nu)\right|^2$ is a function of $\mu$. Analysing the behaviour of curves presented in figures 1 and 2 one can deduce that the shielding of a nuclear charge by a bound electron significantly enhances the amplitude, but insignificantly changes the phase of wavefunction $\psi_{\vec{k}}^{(0)}$. This can clearly be seen on graphs of the real and imaginary parts of wavefunctions (not shown). The same tendency is observable for other energies of a colliding electron.

2.2 Discrete spectrum wavefunctions

The wavefunctions of two bound electrons in $HeI$ can be represented as these symmetric and antisymmetric functions

$$\Psi_f^{(\pm)} = \frac{1}{\sqrt{2}}\left(\psi_1(\vec{r}_1)\psi_2(\vec{r}_2) \pm \psi_1(\vec{r}_2)\psi_2(\vec{r}_1)\right). \tag{11}$$

Here, $\psi_1(\vec{r}_1)$ and $\psi_2(\vec{r}_2)$ are the one-electron wavefunctions, signs (+) and (−) correspond to singlet ($S=0$) and triplet ($S=1$) states, respectively.

The idea of finding $\psi_1(\vec{r}_1)$ and $\psi_2(\vec{r}_2)$ consists in regarding each electron in $HeI$ being in the "self-consistent field" that is created by the nucleus together with the other electron (Hartree 1928). To find wavefunctions $\psi_1(\vec{r}_1)$ and $\psi_2(\vec{r}_2)$ Fock (1930) employed the variational principle. As a result he obtained the system of two coupled integro-differential equations. The derived equations can be solved only numerically. The obtained functions, of course, are not the exact functions but they are the best one-electron wavefunctions obtained in the one-configuration approximation.

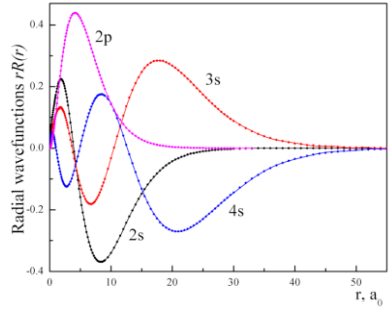

**Fig. 3.** The normalized wavefunctions of an excited electron of helium obtained in the Hartree-Fock approximation; $a_0 = \hbar^2 / m_e e^2 = 0.529 \times 10^{-8}$ cm is the first Bohr radius of hydrogen.

In Fig. 3 are depicted the wavefunctions of an excited electron of $HeI$ obtained in the Hartree-Fock approximation. The electrons are in configuration $(1s)(nl)$ with the total spin $S=0$. For radial wavefunctions $U_{nl} = rR_{nl}(r)$ the numerical data are taken from the book of Bratsev (1979). The wavefunctions obtained with cubic spline interpolation of the numerical functions are shown in solid curves in Fig. 3. At $r \gg 1$ the wave functions are represented as $A\exp(-\gamma r)$ with $A$ determined by fitting this function to the Hartree-Fock wavefunction, and $\gamma = (-2\varepsilon_{nl})^{1/2}$ where $\varepsilon_{nl}$ is the energy of an excited electron. The wavefunctions corresponding to the triplet state have a similar shape but differ in magnitude from functions depicted in Fig. 3.

## 3. TRANSITION PROBABILITIES

In the dipole approximation the probability of a radiative transition is defined as (Heitler 1954)

$$W_{i,f} = \frac{4\omega_{if}^3}{3c^3}\left|\left\langle \Psi_f \left| \vec{d}_1 + \vec{d}_2 \right| \Psi_i \right\rangle\right|^2 = \frac{4\omega_{if}^3}{3c^3}\left|\left\langle \psi_{nlm} \left| \vec{d} \right| \psi_{\vec{k}} \right\rangle\right|^2. \qquad (12)$$

Here, $\omega_{if}$ is the frequency of an emitted photon, $c$ is the speed of light, and $\vec{d} = -(\vec{i}x + \vec{j}y + \vec{k}z)$ is the operator of electric-dipole strength. For convenience, we calculate the matrix elements of operators $d^{(\pm)} = -(x \pm iy)$ and $d^{(z)} = -z$. In parabolic coordinates these operators read $d^{(z)} = -(\mu - \nu)/2$ and $d^{(\pm)} = -\sqrt{\mu\nu}e^{\pm i\varphi}$.

To begin, we calculate the free-bound transition probability using the derived wavefunctions. The non-trivial matrix elements are expressible as

$$d_{if}^{(z)} = -\frac{\sqrt{\pi}}{4}\sqrt{2l+1}\int_0^\infty\int_0^\infty U_{nl}(\mu,\nu)P_l\left(\frac{\mu-\nu}{\mu+\nu}\right)f_{\vec{k}}(\mu,\nu)(\mu-\nu)d\mu d\nu, \qquad (13)$$

$$d_{if}^{(\pm)} = -\frac{\sqrt{\pi}}{2}\sqrt{\frac{(2l+1)(l-1)!}{(l+1)!}}\int_0^\infty\int_0^\infty U_{nl}(\mu,\nu)P_l^1\left(\frac{\mu-\nu}{\mu+\nu}\right)f_{\vec{k}}(\mu,\nu)\sqrt{\mu\nu}d\mu d\nu, \qquad (14)$$

in which $U_{nl}$ is the radial wavefunction of a bound active electron, $f_{\vec{k}} = f_{\vec{k}}^{(0)} + f_{\vec{k}}^{(1)}$ is the wavefunction of a colliding electron and $P_l^m$ is a Legendre polynomial.

In Figs. 4-7 are shown transition probabilities $W_{i,nlm}$ into $2s$, $3s$, $4s$ and $2p_0$-states of $HeI$ as functions of redshift $z$. Probabilities are calculated for singlet states by making use equation (13). The solid curves correspond to the case when the partial shielding of a nuclear charge by a bound electron is taken into account ($\upsilon(r) \neq 0$); the dotted curves exhibit the results of calculations when $\upsilon(r) \equiv 0$. An excited electron in $HeI$ is described by the Hartree-Fock wavefunction (blue curves) or by the hydrogen wavefunctions (red curves).

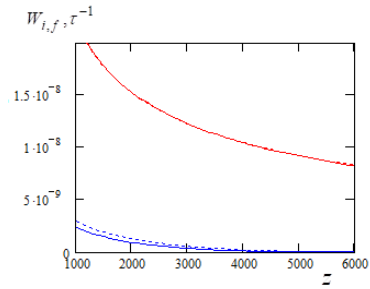

**Fig. 4.** Free-bound transition probability into the $2s$ -state of $HeI$. Solid curves – $\upsilon \neq 0$; dotted curves – $\upsilon \equiv 0$. Blue curves – the Hartree-Fock wavefunctions and red curves – the hydrogen wavefunctions are used for the description of a bound active electron. Solid and dotted red curves are so close that they are almost indistinguishable; the atomic unit of time $\tau = \hbar^3 / m_e e^4 = 2.419 \times 10^{-17} s$.

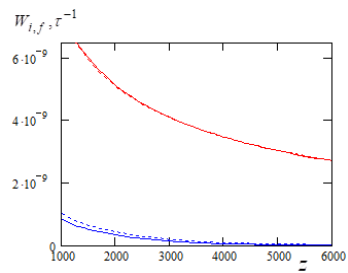

**Fig. 5.** As in Fig. 4 but for the $3s$ -state of $HeI$.

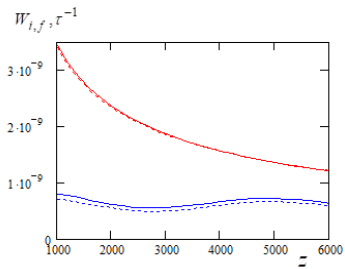

**Fig. 6.** As in Fig. 4 but for the $4s$ -state of $HeI$.

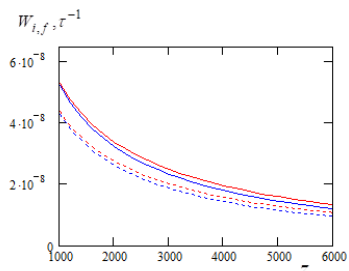

**Fig. 7.** As in Fig. 4 but for the $2p_0$ -state of $HeI$.

Figs. 4-6 demonstrate that transition probabilities $W_{i,2s}$ and $W_{i,3s}$ monotonically decrease when $z$ increases, whereas probability $W_{i,4s}$ has a minimum and a maximum. These minimum and maximum disappear when a colliding electron is described by the hydrogen wavefunction, i.e. when a colliding electron experiences the field of $HeII$ as the purely Coulomb field with charge $Z=1$. As to probability $W_{i,2p_0}$, Fig. 7 shows that $W_{i,2p_0}$ monotonically decreases when $z$ increases. The similar behaviour is observable for $W_{i,np_0}$ with $n>2$.

An interesting result is that free-bound transition probabilities $W_{i,ns}$ weakly depend on the choice of a wavefunction of a colliding electron, whereas strongly depend on the choice of a wavefunction of a bound excited electron. As for probabilities $W_{i,np_0}$ both the shielding effect and the choice of wavefunctions of $HeI$ are insignificant.

The probability of bound–bound radiative transitions in $HeI$ is defined with equation (12) in which

$$\vec{d}_{n'l'm',nlm} = \langle \psi_{n'l'm'} | \vec{d} | \psi_{nlm} \rangle. \tag{15}$$

Here $\psi_{nlm}$ and $\psi_{n'l'm'}$ are wavefunctions of a bound active electron in the initial and final states, respectively. Matrix elements (15) can be readily calculated in spherical polar coordinates by employing the wavefunctions derived in Section 2.

Equations (12)-(15) allow us to calculate the total probability as a function of $z$, which is a product of free-bound and bound-bound transition probabilities. Depending on in which excited state a colliding electron transits, the calculation of one free–bound transition probability $W_{i,nlm}$ takes from several to several tens of seconds on a standard computer.

## 4. CONCLUSIONS

In this paper, we have suggested a rapid and relatively simple scheme of calculation applicable to cosmological recombination of $HeI$. Our approach is based on reducing the two-electron problem to the one-electron treatment. This allows us to represent the wavefunctions of an active electron involved in recombination in a closed algebraic form. The wavefunction of a colliding electron, we find in the Coulomb-Born approximation by making use the Coulomb Green's function defined in an integral form. The bound excited electron in $HeI$, we describe with wavefunctions obtained in the Hartree-Fock approximation. The elaborated scheme of calculation enables us to determine the probability of free-bound radiative transition, in principle, into an arbitrary state of $HeI$.

We have calculated the transition probabilities from the initial continuous spectral state into the low-lying excited states ($n \leq 4$, $l = 0,1$) of $HeI$ as functions of redshift $z$. Furthermore, we have investigated the influence on the transition probability of a shielding of a nuclear charge by the bound electron. An important result is that the transition probabilities weakly depend on to what extent a field experienced by a colliding electron deviates from the Coulomb field with charge $Z=1$. Based on the obtained results, we can state that the assumption that a colliding electron moves in the purely Coulomb field with charge $Z=1$ is an adequate approximation in $HeI$ recombination calculations. This statement allows us to use the Coulomb continuous spectrum wavefunction for the description of a colliding electron. Application of the Coulomb wavefunction substantially simplifies the problem and reduces the time of calculation.

**APPENDIX A**

Here, we evaluate function $Q^{(\pm)}(\vec{r})$ that is defined with expression (5). Wavefunction $\psi^{(0)}(\vec{r})$, we represent as (Landau & Lifshitz 1977)

$$\psi^{(0)}(\vec{r}) = \frac{1}{2k}\sum_{l=0}^{\infty} i^l (2l+1)\exp(i\delta_l) P_l(\cos\vartheta) R_{kl}(r), \qquad (A1)$$

in which $R_{kl}(r)$ are the Coulomb radial functions, $\delta_l$ are the phase shifts of $R_{kl}(r)$ and $P_l(\cos\vartheta)$ are the Legendre polynomials. Making the appropriate calculations, we obtain that

$$Q^{(\pm)}(\vec{r}) = \pm 2 \frac{e^{\pi/k}}{\pi} \sum_{l=0}^{\infty} \frac{i^l e^{i\delta_l}(2k)^l \left|\Gamma(l+1-2i/k)\right|}{(2l+1)!}$$
$$\cdot P_l(\cos\vartheta)\left(C_l(r)r^{-l-1} + D_l(r)r^l\right)\psi_{1s}(\vec{r}), \qquad (A2)$$

in which

$$C_l = \int_0^r e^{-(2+ik)r'} F\left(\frac{2i}{k}+l+1, 2l+2, 2ikr'\right) r'^{2l+2} dr', \qquad (A3)$$

$$D_l = \int_r^\infty e^{-(2+ik)r'} F\left(\frac{2i}{k}+l+1, 2l+2, 2ikr'\right) r' dr', \qquad (A4)$$

When $r$ tends to infinity integral (A3) becomes solvable (Gradshtein & Ryzhik 1980)

$$\int_0^\infty e^{-(2+ik)r'} F\left(\frac{2i}{k}+l+1, 2l+2, 2ikr'\right) r'^{2l+2} dr'$$
$$= \frac{(2l+2)!}{(2+ik)^{2l+3}} F\left(\frac{2i}{k}+l+1, 2l+3, 2l+2, 2ik/(2+ik)\right), \quad (A5)$$

where $F(a,b,c,x)$ is a hypergeometric function. As to integral (A4), it tends to zero when $r$ tends to infinity. We thus obtain that $Q^{(\pm)}(\vec{r})$ exponentially decreases, whereas $V(r)$ decreases as $1/r$ when $r \to \infty$.

**APPENDIX B**

Here, we define functions that are appeared in equation (10):

$$A^{(n)}(s,\mu) = \int_0^\infty e^{-a_\mu \mu'} J_0\left(b\sqrt{\mu\mu'}\right) \mu'^n d\mu', \quad (B1)$$

$$B^{(n)}(s,\nu) = \int_0^\infty e^{-a_\nu \nu'} J_0\left(-b\sqrt{\nu\nu'}\right) F\left(i/k, 1, ik\nu'\right) \nu'^n d\nu'. \quad (B2)$$

The integral (B1) is analytically solvable (Gradshtein & Ryzhik 1980)

$$A^{(n)}(s,\mu) = \frac{\Gamma(n+1)}{a_\mu^{n+1}} F\left(n+1, 1, -\frac{b^2}{4a_\mu}\mu\right). \quad (B3)$$

Expanding the confluent hypergeometric function over an argument

$$F(i/k, 1, x) = 1 + \frac{(i/k)}{1}\frac{x}{1!} + \frac{(i/k)(i/k+1)}{1 \cdot 2}\frac{x^2}{2!} + \cdots, \quad (B4)$$

(B2) can be reduced to a sum of analytically solvable integrals

$$B^{(n)}(s,\nu) = \sum_{l=0}^\infty (ik)^l d_l \int_0^\infty e^{-a_\nu \nu'} J_0\left(-b\sqrt{\nu\nu'}\right) \nu'^{n+l} d\nu'$$
$$= \sum_{l=0}^\infty (ik)^l d_l \frac{\Gamma(n+l+1)}{a_\nu^{n+l+1}} F\left(n+l+1, 1, -\frac{b^2}{4a_\nu}\nu\right). \quad (B5)$$

Here $d_0 = 1$, $d_1 = i/k$, $d_2 = (i/k)(i/k+1)/(2!)^2$, $d_3 = (i/k)(i/k+1)(i/k+2)/(3!)^2$ and so on. Calculation shows that convergence is rapid in (B5).